\documentclass[aip,reprint]{revtex4-1}

\draft 

\usepackage{mathptmx}
\usepackage[T1]{fontenc}
\usepackage{graphicx} 
\usepackage{amssymb}
\usepackage{etoolbox}
\usepackage{bm}

\usepackage{xcolor}

\begin{document}


\title{Dynamical Data for More Efficient and Generalizable Learning: A Case Study in Disordered Elastic Networks } 




\author{Salman N. Salman}
\thanks{These authors contributed equally to this work.}
\affiliation{The Wolfson Department of Chemical Engineering, Technion -- Israel
Institute of Technology, Haifa 32000, Israel}

\author{Sergey A. Shteingolts}
\thanks{These authors contributed equally to this work.}
\affiliation{The Wolfson Department of Chemical Engineering, Technion -- Israel
Institute of Technology, Haifa 32000, Israel}

\author{Ron Levie}
\affiliation{Faculty of Mathematics, Technion -- Israel
Institute of Technology, Haifa 32000, Israel}

\author{Dan Mendels}
\email[E-mail: ]{danmendels@technion.ac.il}
\affiliation{The Wolfson Department of Chemical Engineering, Technion -- Israel
Institute of Technology, Haifa 32000, Israel}


\begin{abstract}
Machine learning models often require large datasets and struggle to generalize beyond their training distribution. These limitations pose significant challenges in scientific and engineering contexts, where generating exhaustive datasets is often impractical and the goal is frequently to discover novel solutions outside the training domain. In this work, we explore the use of dynamical data through a graph neural network-based simulator to enable efficient system-to-property learning and out-of-distribution prediction in the context of uniaxial compression of two-dimensional disordered elastic networks. We find that the simulator can learn the underlying physical dynamics from a small number of training examples and accurately reproduce the temporal evolution of unseen networks. Notably, the simulator is able to accurately predict emergent properties such as the Poisson's ratio and its dependence on strain, even though it was not explicitly trained for this task. In addition, it generalizes well across variations in system temperature, strain amplitude, and most significantly, Poisson's ratios beyond the training range. These findings suggest that using dynamical data to train machine learning models can support more data efficient and generalizable approaches for materials and molecular design, especially in data-scarce settings.
\end{abstract}


\maketitle 

\section{Introduction}
\label{intro}
The use of artificial intelligence (AI) in science and engineering has grown rapidly in recent years, driven by advances in hardware, the increasing availability of large datasets, and algorithmic breakthroughs.\cite{Mowbray2021MachineReview,Liu2021MachineReview,Wang2023ScientificIntelligence,Notin2024MachineDesign}
AI and machine learning (ML) have proven effective across a range of tasks, including protein structure prediction, \cite{Jumper2021HighlyAlphaFold,Krishna2024GeneralizedAll-Atom} force field calculation in molecular simulations,\cite{Behler2007GeneralizedSurfaces,Piaggi2024AFeldspar,Zhang2022AInteractions,Bartok2018MachineSilicon,Cheng2020EvidenceHydrogen} enhanced sampling,\cite{Zhang2024Descriptor-FreeNetworks, Bonati2021DeepSampling,Bonati2023AMlcolvar, Mendels2018FoldingAnalysis,Mendels2018CollectiveFluctuations,Piccini2018MetadynamicsChemistry,Rizzi2019BlindAnalysis,Noe2019BoltzmannLearning} structure–property mapping for molecular and materials systems, \cite{Hollingsworth2018MolecularAll,Chong2024AdvancesTechniques,Zhang2023ArtificialSystems, Avery2019PredictingSearch,Ojih2023ScreeningNetwork} and generative modeling for inverse design. \cite{Yao2021InverseModels,Liu2018GenerativeMetasurfaces,Bastek2023InverseModels}

Despite their success, ML-based methods often require large training datasets, which are frequently unavailable or prohibitively expensive to obtain. \cite{Mendels2022CollectiveEvents,Alzubaidi2023AApplications, Chang2022TowardsFramework} Additionally, ML-based models typically struggle to extrapolate beyond their training domain, \cite{Meredig2018CanDiscovery, Li2023AProperties, Omee2024Structure-basedStudy, Kauwe2020CanMaterials} making them inefficient for out-of-distribution (OOD) prediction and generation tasks. This poses a major challenge for molecular and materials engineering, where the discovery of novel systems with desired properties is a key objective.

One promising approach to mitigate these challenges is to use dynamical data from the domain of interest for model training. Such data can often be obtained from simulations and, in some cases, from experiments where dynamical behavior is measurable.\cite{Colen2021MachineHydrodynamics} The central idea is to train ML models to recognize recurring patterns in system dynamics, allowing them to capture underlying physical principles that go beyond static structural information or global descriptors. \cite{Cranmer2020DiscoveringBiases} Thus, by capturing underlying dynamical regularities, such models can achieve substantial efficiency gains - substantially reducing the number of systems needed for training. In turn, this can significantly lower the overall demand for training data from resource-intensive sources such as experiments, \textit{ab initio} calculations, or  simulations involving extensive sampling over long timescales

To this end, neural network (NN) simulators \cite{Wiewel2019LatentFlow,Ladicky2015Data-drivenForests,Kochkov2021MachineDynamics,Sanchez-Gonzalez2020LearningNetworks,Pfaff2020LearningNetworks,Liu2023LearningSimulator, Kumar2023GNS:Modeling} that learn to predict system dynamics offer a promising approach. Unlike models trained to predict specific properties, NN simulators can be leveraged to infer a wide range of system properties without requiring explicit supervision for each property. In addition, they provide predictions for the temporal behavior of a system rather than only its static properties.  Finally, their end-to-end differentiability makes them well suited for force field development and inverse design tasks,\cite{Allen2022PhysicalSimulators} enabling optimization for both target properties and dynamical behavior.

Neural network-based simulators have attracted increasing attention in recent years, primarily due to their potential to efficiently capture system dynamics and substantially reduce the computational cost associated with simulating complex systems. \cite{Li2022GraphDynamics,Pfaff2020LearningNetworks} Unlike traditional simulators that rely on domain-specific equations of motions or those that use neural networks to compute force fields, NN-based simulators hinge on learning the dynamics of a given domain of systems directly. This enables them to predict system behavior using significantly larger time steps than conventional methods (a.k.a. leap-frogging), often achieving time steps orders of magnitude larger than the latter.\cite{Klein2023Timewarp:Dynamics, Liu2023LearningSimulator}

\begin{figure}
    \centering
    \includegraphics[scale=1]{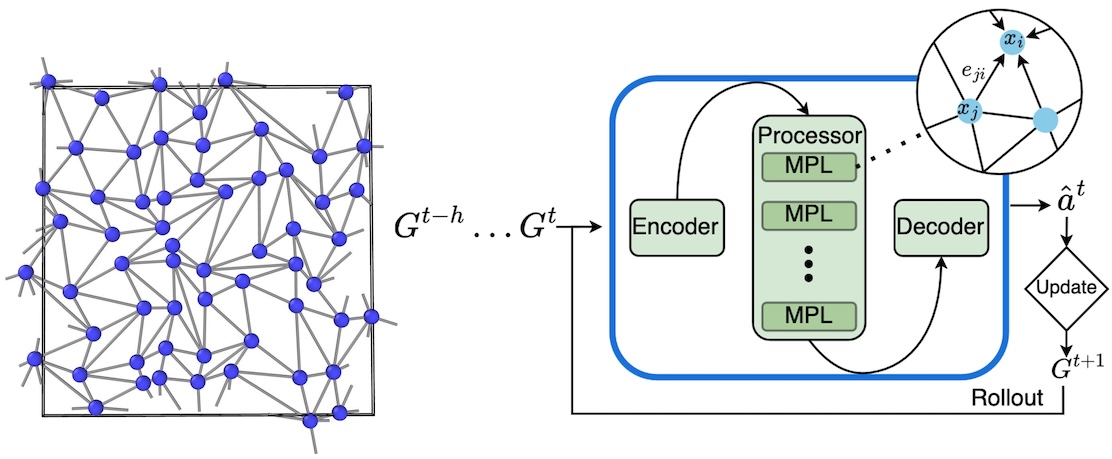}
    \caption{Left: Illustration of a two-dimensional disordered elastic network with periodic boundary conditions, where particles represent point masses and edges correspond to harmonic bonds. Right: Architecture of the simulator model used in this work. MPL denotes a message passing layer; layers are applied sequentially to the processor’s input.}
    \label{fig:architecture}
\end{figure}

In this work, we investigate the predictive capabilities of a graph neural network (GNN)-based simulator, developed based on a recently introduced framework,\cite{Sanchez-Gonzalez2020LearningNetworks,Pfaff2020LearningNetworks} for modeling the dynamics of uniaxial compression in two-dimensional disordered elastic networks (DENs) (see Fig. \ref{fig:architecture} for illustration). DENs are highly tunable systems capable of exhibiting a broad range of mechanical behaviors and properties. They have been employed to model diverse systems, including amorphous solids,\cite{Lerner2014BreakdownSolids} mechanical metamaterials,\cite{Reid2018} biological tissue,\cite{Manning2024RigidityNetworks} functionality design,\cite{Berneman2024DesigningNetworks, Mendels2023SystematicTailoring, Shen2024AnDensity} learning physical systems,\cite{Stern2021SupervisedMachines,Stern2023LearningSystems} and proteins.\cite{Bahar2010GlobalFunction} We find that our GNN simulator effectively captures the compression dynamics of the DENs, successfully predicts their Poisson's ratio as an emergent property, and effectively reproduces its strain-dependent behavior. Furthermore, the simulator demonstrates notable data efficiency, requiring training on only a small number of systems to achieve accurate predictions across a wide range of systems in the test set. Finally, it exhibits promising OOD generalization capabilities, particularly in predicting the dynamics of systems with Poisson's ratios well outside the range encountered during training. Such generalization may be especially valuable for design applications, as it could enable the exploration and generation of novel systems with mechanical functionalities and behaviors not explicitly represented in the training data.

\section{Methods}
\label{methods}
\subsection*{The Simulator}
The simulator’s objective is to predict the next time step in a DEN's compression trajectory based on the system’s previous configuration/s, thereby enabling the auto-regressive generation of full trajectories from a given set of initial conditions. There is some flexibility in defining the output of the predictor. For instance, the model can be set to predict the next-step positions of the system's particles, or alternatively, higher-order derivatives of the positions, such as acceleration. We have found that  predicting acceleration yielded the best model performance, which agrees with previous results in the literature.\cite{Sanchez-Gonzalez2020LearningNetworks} The predicted acceleration is then used in a forward Euler integration ($\Delta t = 1$) scheme, $\textbf{r}^{t+1}_{i} = {\hat{\textbf{a}}}^{t}_i + 2\textbf{r}^{t}_i - \textbf{r}^{t-1}_i$ to compute the next-step positions $\textbf{r}^{t+1}_i$. Namely, starting from a system configuration at timestep $t$, denoted $G^t$, the model predicts the corresponding particle accelerations $\hat{\textbf{a}}^t$, which are then used to construct the next configuration $G^{t+1}$. Next, we recompute the edge vectors (which represent in our case interparticle bonds) and their corresponding lengths based on the inferred $\textbf{r}^{t+1}_i$. For more details on the data representation see ``Data Representation.'' 
 This process is repeated auto-regressively to generate a full rollout  $\{ G^t, G^{t+1}, ..., G^{t+l_{rollout}}\}$, where $l_{rollout}$ is the rollout length.

\subsection*{Elastic networks}
Details\cite{Parrinello1981PolymorphicMethod,Schneider1978Molecular-dynamicsTransitions} on the DEN generation and compression protocol can be found in Sec. I of the supplementary material. In short, DENs were generated in a two-step process\cite{Rocks2017DesigningNetworks} using a standard jamming algorithm. \cite{Liu2010TheSolid} Harmonic bonds between the point-sized particles were assigned an elastic energy in the form:
\[E(r_{ij}) = K_b (l - l_{r})^2, \]
where $K_b$ is the bond coefficient, $l_r$ is the bond rest length, and $l$ is the distance between the two particles. For each bond, the value of $K_b$ was set to $1/l_{r}$. For each pair of bonds sharing one common particle an angle with an harmonic potential was assigned, with the energy in the form:
\[E(\theta_{km}) = K_{ang} (\theta_{km} - \theta_{km}^0)^2, \]
where $K_{ang}$ is the angle coefficients for angle $\theta_{km}$ between the pair of edges $k$ and $m$, and $\theta_{km}^0$ is the rest angle. Throughout this work we use Lennard Jones units (see Sec. I in the supplementary material).

\subsection*{Data Representation}
Given the inherently graph-based structure of our physical systems, employing a graph neural network is a natural modeling choice,  an approach that has also been widely adopted in the broader context of interacting particle systems. The state of the system at each timestep is thus encoded into a graph $G$, containing a set of nodes $X$ of size $N$ and edges $E$ of size $M$. Each graph node corresponds to an elastic network particles and has an associated one-dimensional feature vector $\textbf{x}_i$, which contains information about the node velocity, calculated as $\textbf{v}^t_i=\textbf{r}^t_i - \textbf{r}^{t-1}_i$, where $\textbf{r}^{t}_i$ is the $i$'s node position at time $t$ and $\textbf{r}^{t-1}_i$ its position at time $t-1$. More generally, node velocity vectors can be of the dimension $\mathbb{R}^{2h}$, with $h$ ranging from 1 to 3, referring to the number of history steps (previous system configurations) included in the model input. Similarly, the edge features $\textbf{e}$ in the graph correspond to the DEN's harmonic bonds and represent bonded interactions. For each graph edge $\textbf{e}_{ij}$ representing a bond between particles $i$ and $j$, the corresponding features are composed of the edge vector $\textbf{r}_i - \textbf{r}_j$, the length $l$ of the edge vector and a stiffness factor that determines the strength of the bond, defined as $1/l_{r}$ at rest length. This data representation is invariant to  translations. Since these symmetries are built into  the architecture by design, the model does not need to waste training data and parameters to learn them.

\subsection*{Model architecture}
The model follows an encoder-processor-decoder architecture.\cite{Hamrick2018RelationalMachines,Pfaff2020LearningNetworks} The encoder  maps the input node and edge features of a given graph $G$ into latent vectors of size 128, using learnable multi layer perceptrons (MLPs) $\varepsilon^x$ and $\varepsilon^e$ for the node $\{\textbf{x}_i\} \in \mathbb{R}^{N \times 2h}$ and edge $\{\textbf{e}_i\} \in \mathbb{R}^{M \times 4}$ features, respectively. 
The processor consists of $k$ identical message passing (MP) layers, each with its own set of learnable parameters. These layers are applied sequentially, with each layer operating on the output of the previous one. Node and edge embeddings are updated according to the following relation: 
\[\textbf{x}_{i}^{(k)} = \psi^{(k)}(\textbf{x}_{i}^{(k-1)}, \sum_{j\in\mathcal{N}(i)}\phi^{(k)}(\textbf{x}_{i}^{(k-1)}, \textbf{x}_{j}^{(k-1)},\textbf{e}_{ij})\]
where $\psi, \phi$ are learnable MLPs, $\textbf{x}_{i}^{(k-1)}$ denotes the embedding of  node $i$ in the $(k-1)$th layer of the processor, and summation is taken over all nodes $j$ which are bonded to $i$, i.e., which belong to the \emph{neighborhood} $\mathcal{N}(i)$ of $i$. We experimented with replacing the MP layers with attention layers,  \cite{Shi2021MaskedClassification} but observed no significant difference in performance.

The output of the processor is mapped by a decoder to a two-dimensional output $\{\hat{\textbf{a}}_i\} \in \mathbb{R}^{N \times 2}$, representing accelerations of the nodes, using an additional MLP denoted  $\delta^x$. All  MLPs in the architecture, including  $\varepsilon^x, \varepsilon^e, \psi, \phi$, and $\delta^x$ were implemented as two-layer feed-forward networks with hidden dimensions of 128 and ReLU activations.

\subsection*{Implementation and training details}
The model was implemented with PyTorch Geometric 2.5.2,\cite{Fey2019FastGeometric} and PyTorch 2.2.2.\cite{Paszke2019PyTorch:Systems} 
To generate the training data we used uniaxial compression trajectories consisting of systems with varying Poisson's ratios $\nu \in [-0.6,0.4]$ values, except for the models in Fig. \ref{fig:generalization-comparison}. In total, 3500 trajectories were generated, although we found only a small   fraction of that was necessary to achieve high accuracy (see Fig. \ref{fig:data_scaling}). For testing, we used 200 trajectories with the same range of $\nu$ as the training. Both input (i.e. node and edge features) and output data were online normalized separately during the training to zero-mean and unit-variance. Namely, at each forward step, the data are normalized with respect to all their past values seen by the model up to that point. \cite{Sanchez-Gonzalez2020LearningNetworks, Kumar2023GNS:Modeling} All the learnable weights and biases of the model were randomly initialized from a normal distribution with zero mean and a standard deviation of $1/\sqrt{n}$, where $n$ is the number of weights in the layer. We experimented with different initialization schemes and found that $\mathcal{N}(0,1/\sqrt{n})$ yields a better performance overall than a uniform distribution $\mathcal{U}(-1/\sqrt{n},1/\sqrt{n})$. The model was trained in a supervised manner to minimize the mean squared error (MSE) loss between the predicted acceleration $\hat{a}^{t}$ and the ground truth acceleration $a^{t}$.
\[\mathcal{L} = \frac{1}{N} \sum_{i=1}^{N} ({\hat{\textbf{a}}}^{t} - \textbf{a}^{t})^2\]
To minimize the loss, we used the Adam optimizer \cite{Kingma2014Adam:Optimization} with an exponential learning rate scheduler. We set the learning rate to decay from $10^{-4}$ to $10^{-6}$ over 35,000,000 steps with a decay constant of $0.995$.

\subsection*{Training data generation}
The training data was generated using the LAMMPS molecular dynamics (MD) software package \cite{Thompson2022LAMMPSScales} and consists of DEN compression simulation trajectories. Briefly, each dataset entry corresponds to a compression trajectory comprising a sequence of system configurations sampled from a ground truth simulation at regular timesteps $t_s$, beginning from the initial timestep $t=0$. To facilitate the learning process, the sampling periods were adjusted to maintain consistent per-step node velocities across different system sizes. Specifically, the sampling period $\xi$ for each simulation was determined according to:  

\[\xi = \frac{\Delta l_{t}}{s_{r}\; l^{0}}, \] 
where $\Delta l_{t}$ denotes the constant change in box size along the compression axis between consecutive snapshots across all simulations, $s_{r}$ is the strain rate, and $l^0$ is the initial box length along the compression axis. 

\section{Results and Discussion}
\label{results}
\subsection{Model performance}

The model performance was evaluated based on its ability to replicate the system's dynamics. To this end, we defined three metrics: the MSE on the next-step node positions to verify prediction accuracy, the coefficient of determination $R^2$ to measure the agreement between predicted $\hat{\nu}$ and ground truth $\nu$ values, and the absolute error $\epsilon_{\nu}$ in prediction of these values to quantify the model's generalization capabilities on unseen dynamics.

\begin{figure*}
    \centering
    \includegraphics{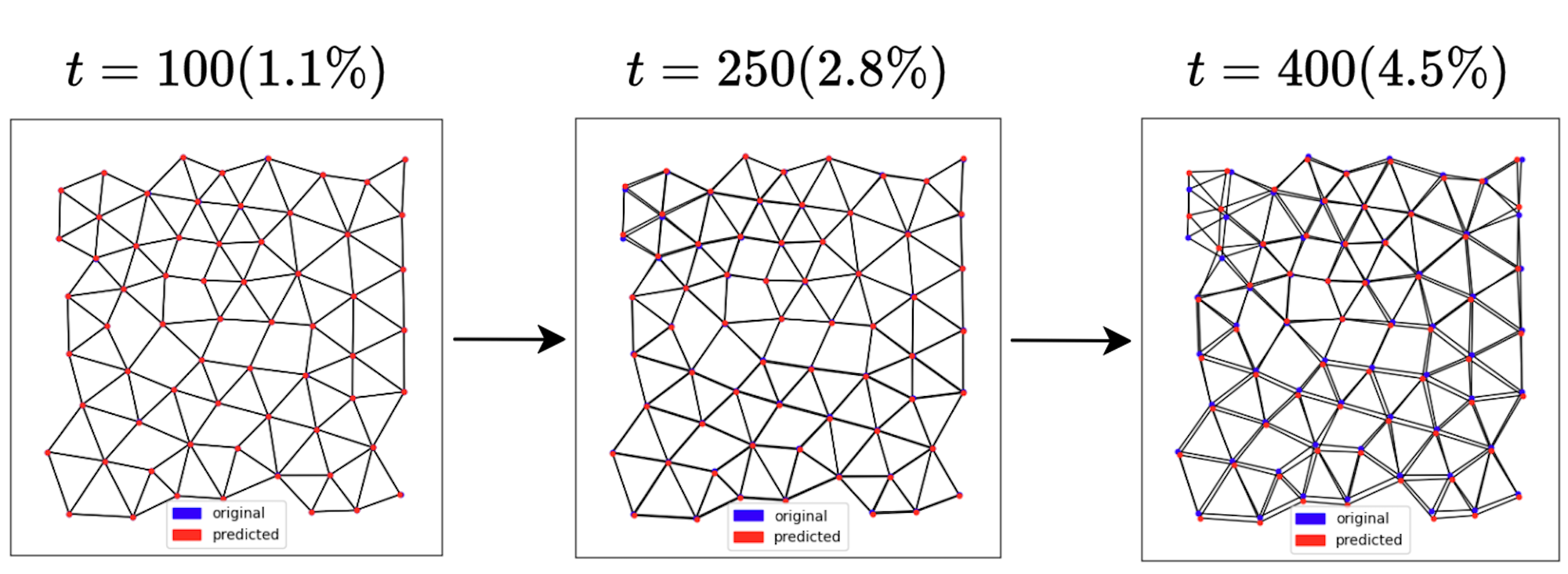}
    \caption{Predicted network configurations (red) from a compression rollout overlaid on top of the ground truth ones (blue) at specified timesteps $t$ and corresponding strains.}
    \label{fig:evolvement}
\end{figure*}
\begin{figure}
    \centering
    \includegraphics{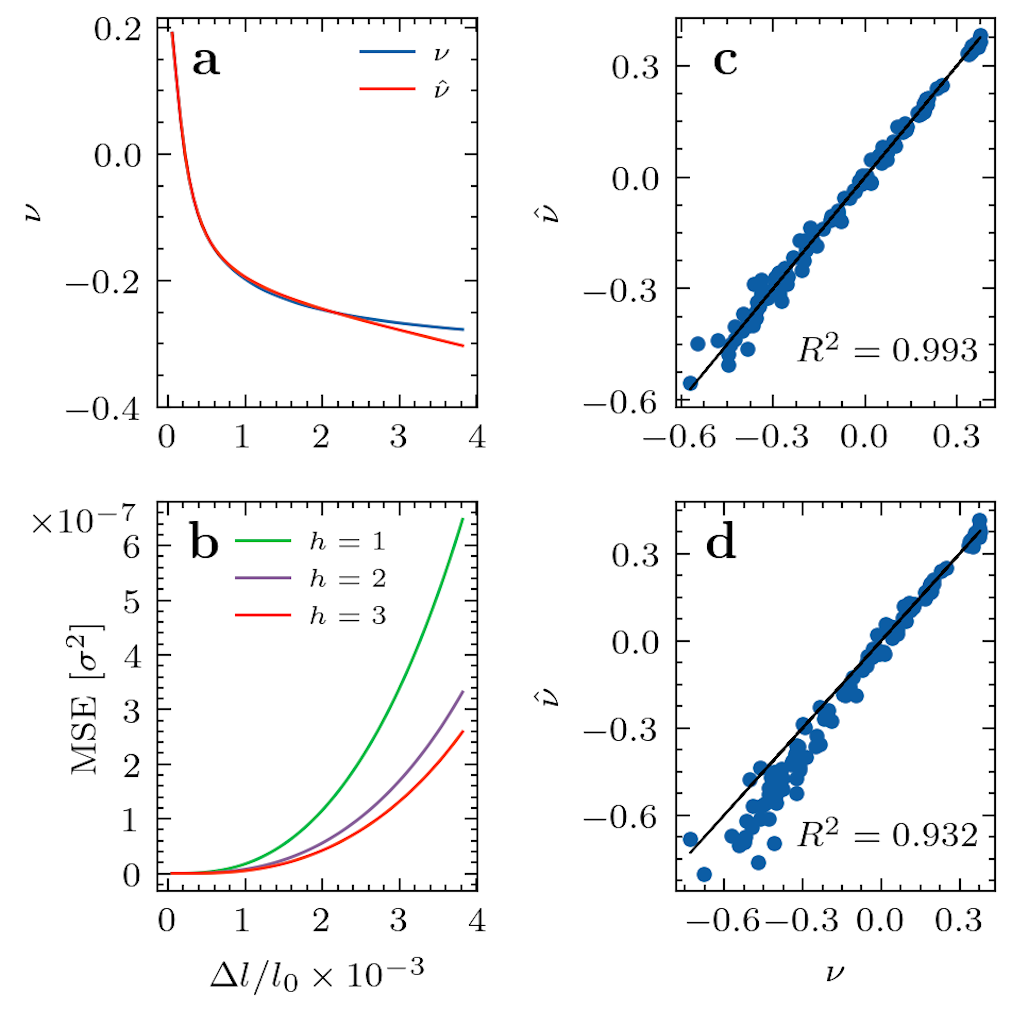}
    \caption{Predicted and ground truth Poisson's ratio $\nu$ (a) and position MSE (b) as functions of strain. Green, magenta and red lines correspond to 2, 3 and 4 model input configurations, respectively. Parity plots of the predicted Poisson's ratio $\hat{\nu}$ vs the ground truth values $\nu$ for 20 step (c) and 100 step (d) rollouts.}
    \label{fig:preformance}
\end{figure}

We evaluated the performance of our model on strains of up to 5\%. As can be seen from the example in Fig. \ref{fig:evolvement}, even after 400 consecutive predictions, corresponding to a strain of 4.5\%, the final network structure is reproduced quite well, featuring only minor discrepancies. As a quantitative measure of the simulator's predictive capabilities, we focused specifically on its ability to predict Poisson's ratio of previously unseen DENs. To assess this, we compared predicted $\nu$ values with ground truth results after rollouts of $t=20$ and $t=100$ simulator steps (Fig. \ref{fig:preformance}c \& d). Poisson's ratios were estimated using a custom-written function (see Sec. I of the Supplementary Material), which calculates the average node displacement in the $x$ and $y$ directions between $t=0$ and the final rollout step. We found that the simulator successfully reproduces the correct Poisson's ratios, despite not being explicitly trained for this task. Interestingly, the model was also able to effectively capture the dependence of the networks' Poisson's ratio on the applied strain, as shown in Fig. \ref{fig:preformance}a.  

In general, we observed that the simulator was more accurate at replicating non-auxetic ($\nu \geq 0$) compression trajectories, with very minimal errors for both short and long rollouts. Conversely, prediction errors for auxetic ($\nu < 0$) compression were noticeably larger, increasing significantly with longer rollouts (Fig. \ref{fig:preformance}b).

Importantly, we found that our model demonstrates notable data efficiency, achieving high prediction accuracy even when trained on a small number of system trajectories 
(see Fig. \ref{fig:data_scaling}). This stands in contrast to the common case where structure–property ML models require substantially larger datasets.\cite{Dold2023DifferentiableMaterials,Ojih2023ScreeningNetwork, Kauwe2020CanMaterials} These results highlight the potential effectiveness of the proposed approach in improving data efficiency for learning structure–property relationships using ML. Such efficiency can be particularly valuable in domains where data acquisition is prohibitively expensive, including those requiring experiments, \textit{ab initio} simulations, or simulations involving extensive sampling over long timescales.  

Overall, as expected, the position MSE increases with rollout length, rising from approximately $1 \times 10^{-10}$ for single-step predictions to around  $1\times 10^{-5}$ after 100 consecutive steps (Fig. \ref{fig:preformance}b). The model’s prediction accuracy is strongly influenced by the number of history steps it utilizes as features. As shown in Fig. \ref{fig:preformance}b, increasing the number of past velocity inputs significantly improves the model’s prediction accuracy and rollout stability, enabling longer and more reliable prediction chains. At the same time, this benefit exhibits diminishing returns, with no noticeable improvement observed for history lengths greater than $h>3$.

The simulator's accuracy is not significantly affected by the number of MP layers. By testing several values of $k$, we found that increasing the number of MP layers leads only to a marginal increase in performance. Using more than one layer yields diminishing returns while significantly increasing the model complexity and training time.

\subsection{Compression dynamics}
\begin{figure}
    \centering
    \includegraphics{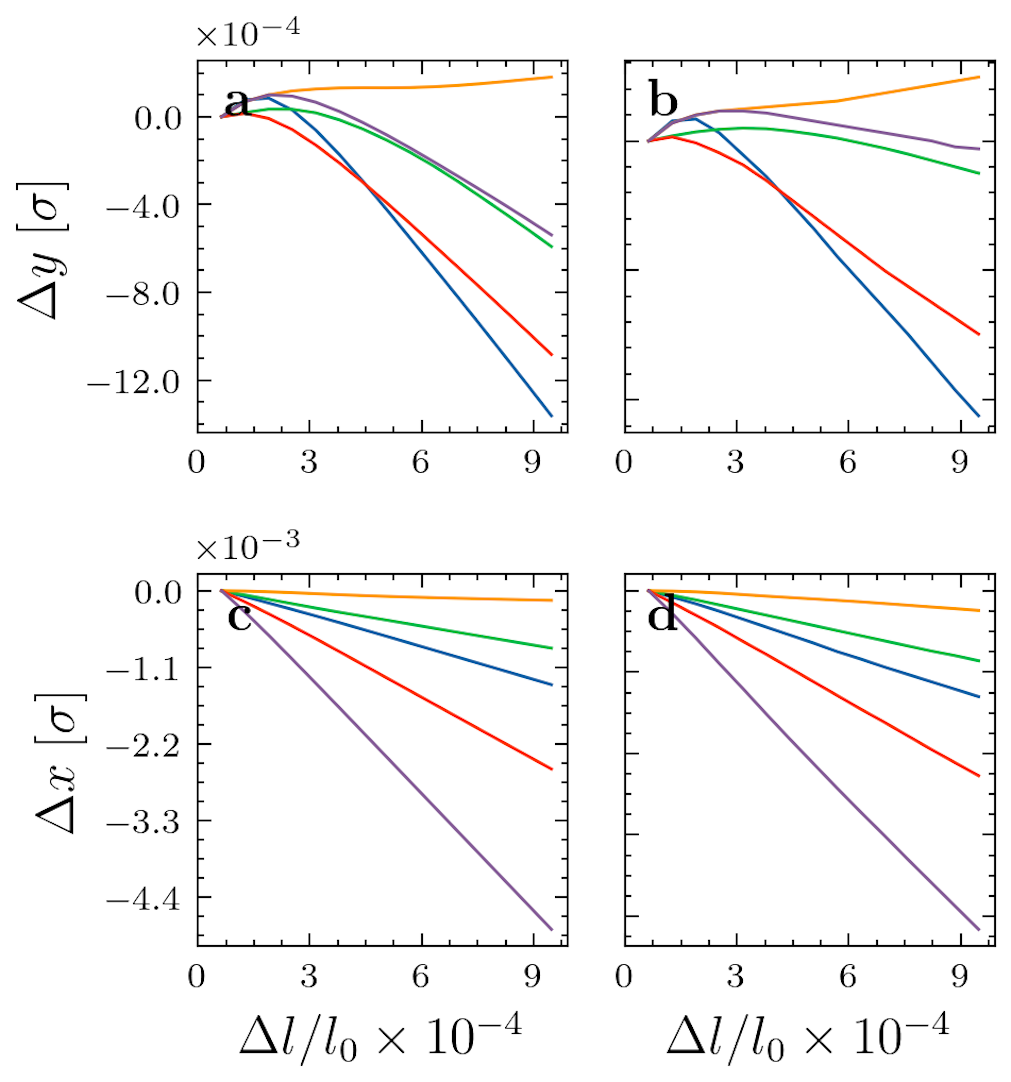}
    \caption{ Individual node trajectory predictions (left) versus ground truth (right) for a compression simulation of an auxetic network with $\nu \approx-0.3$.  In (a, b) the displacement of the $y$ node components and in (c, d) the displacement of the  $x$ node components. The first 15 steps of the rollout are shown.}
    \label{fig:node_trj}
\end{figure}

\begin{figure}
    \centering
    \includegraphics{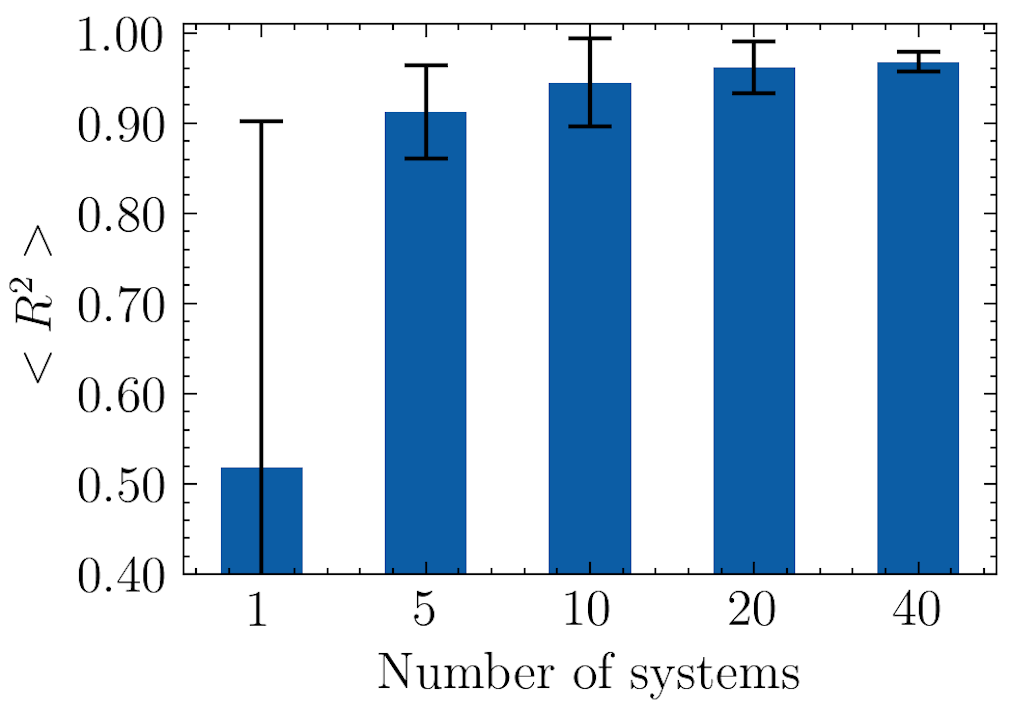}
    \caption{Average Poisson's ratio prediction accuracy after 20 step rollout as a function dataset size. Error bars indicate standard deviation of $R^2$ distribution. ($R^2$ was attained from parity plots such as those shown in Fig. \ref{fig:preformance})}
    \label{fig:data_scaling}
\end{figure}

A closer examination of the simulator’s rollouts reveals that it is capable of qualitatively reproducing the ground truth microscopic individual node trajectories, as illustrated in Fig. \ref{fig:node_trj}. We find to this end, however, that its performance is strongly influenced by the nature of the simulated dynamics, particularly the type of system undergoing compression. In non-auxetic networks, node trajectories tend to exhibit homogeneous and relatively simpler dynamics in both spatial dimensions, $x$ and $y$ (see Sec. II in the supplementary material). In contrast, the $y$-axis trajectories in auxetic networks reveal two distinct regims (Fig. \ref{fig:node_trj}b, see also Fig. 1 in sup). The first regime, occurring at the beginning of the simulation, is marked by complex, non-monotonic motion. This is followed by a second regime characterized by simpler, generally monotonic behavior. To quantify the complexity for a system with $N$ nodes and $L$ snapshots (time window size considered for the calculation), we defined the complexity measure
\[ \alpha = \frac{1}{NL}\sum_{i = 0}^{N} \sum_{t=t_0}^{L} |\textbf{v}_{t+1,i} -\textbf{v}_{t,i}|, \]
where $\textbf{v}_{t,i}$ represents the $i$-th node velocity at time $t$. Larger $\alpha$'s indicate more complex dynamics in the system on average. We have found that for non-auxetic systems, compression is described by a nearly time-independent $\alpha$ value, while for auxetic systems the complexity gradually decreases from a larger $\alpha$ at $t=0$ towards that same value at around $t=10$ (see Fig. 2 in supplementary material). Accordingly, the observed inaccuracies in predicting auxetic compression trajectories can be attributed to the increased complexity and heterogeneity of the dynamics at the beginning of the simulations. In this region, the model tends to perform less accurately, leading to greater error accumulation over the course of the rollout.

\subsection{Beyond Training Distribution Generalization}
\label{generalization}

\begin{figure}
    \centering
    \includegraphics{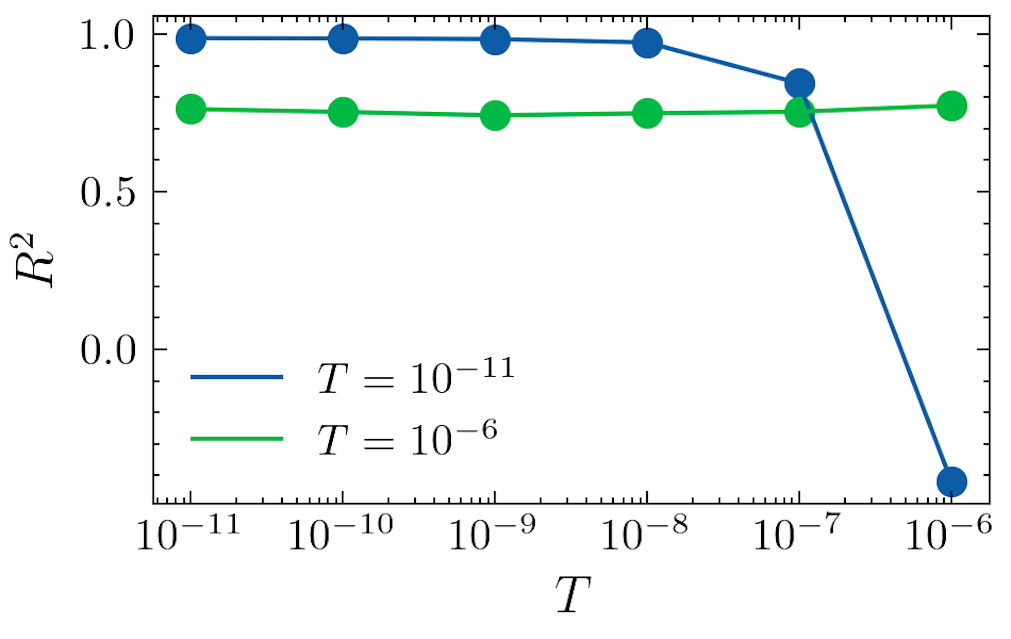}
    \caption{Performance comparison of a model trained with high temperature data ($T=10^{-6}$) and a model trained on low temperature data ($T=10^{-11}$). ($R^2$ was attained from parity plots such as those shown in Fig. \ref{fig:preformance}.)}
    
    \label{fig:temp-gen}
\end{figure}

To evaluate the simulator’s ability to generalize beyond its training distribution, we tested its predictive performance on systems exhibiting both conditions and intrinsic properties not encountered during training. Fig. \ref{fig:temp-gen} illustrates the simulator’s ability to generalize with respect to system temperature. We found that training on low-temperature data (starting at $T = 10^{-11}$) was sufficient for accurate Poisson's ratio predictions up to $T= 10^{-7}$. Moreover, models trained on lower $T$ data generally perform better across the full range of temperatures up to $T= 10^{-7}$. However, this no longer holds for $T\geq10^{-7}$, where the same accuracy level can only be achieved by training in the similar temperature band (see Fig. \ref{fig:temp-gen}). We attribute this to the thermal energy of the system $k_{B}T$ becoming comparable to the typical bond energy. This suggests that the model is capable of learning the relevant dynamics even within a noisy, high-temperature regime. This capability is further illustrated in Fig. 3 in supplementary material, which compares ground truth and predicted node trajectories under high-temperature conditions. 

We were particularly interested in the simulator’s ability to generalize beyond its training distribution of Poisson's ratios, given the relevance of such generalization for inverse design and materials discovery. To evaluate this, we divided the training data into smaller subsets of systems, each containing simulations within a specific $\nu$ range. We then trained separate models on each subset individually and evaluated their performance across the full range of Poisson's ratios, including the range used for training.

\begin{figure}[h!]
    \centering
    \includegraphics{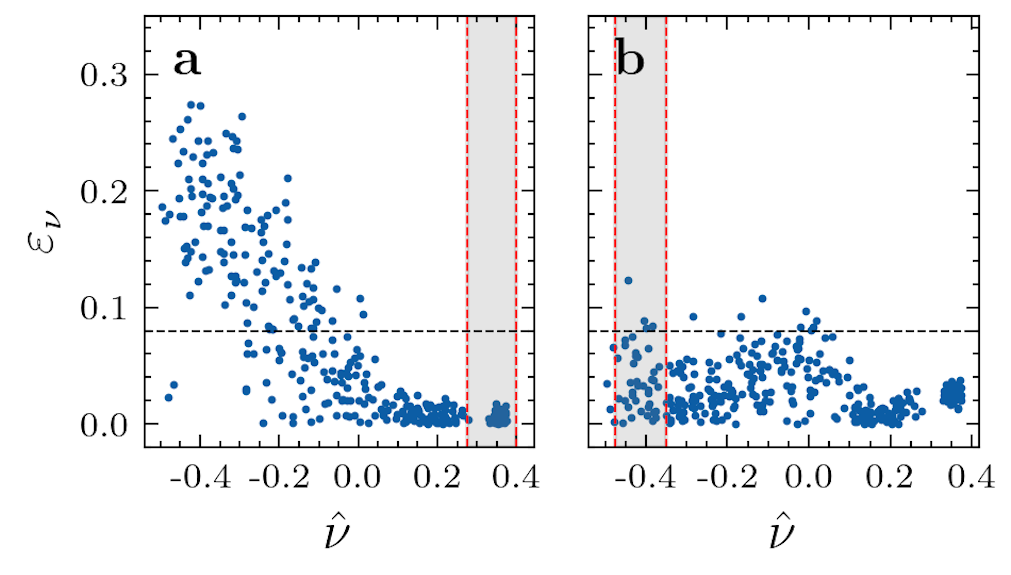}
    \caption{Absolute error of Poisson's ratio prediction for a model trained only on non-auxetic (a) and auxetic (b) data. The gray shaded regions indicate the range of Poisson's ratios  included in the training data.}
    \label{fig:generalization-comparison}
\end{figure}

Fig. \ref{fig:generalization-comparison} features two representative cases: one model trained on data with a high Poisson's ratio range and another on data with a low (highly auxetic) range. As can be seen from the figure, the simulator trained on the highly auxetic data with $\nu \in [-0.6,-0.475]$ demonstrated strong generalization performance across the full range of considered Poisson's ratios. In contrast, the model trained on higher $\nu$ data exhibited good generalization ability, but only down to $\nu \approx0$. This suggests that generalization from more complex and heterogeneous dynamics (high $\alpha$) to simpler, more homogeneous dynamics (low $\alpha$) can be quite effective, likely because the model is exposed to a broader range of patterns in the complex scenario, some of which are also relevant in the simpler case.  Although generalization in the opposite direction is more constrained, a meaningful level of out-of-distribution generalization is still achieved. 

Finally, we also found that the model could achieve good accuracy on networks up to approximately twice the size (in number of nodes) of those it was trained on. Additionally, we found that limiting training to only the high-$\alpha$ region, for example the first 20 steps of the trajectories, did not degrade performance. The model was still able to accurately generate longer rollouts, highlighting its generalization ability with respect to strain amplitude. Notably, it could predict compressions corresponding to strains up to 20 times larger than those seen during training. 

\section{Conclusion}
\label{conclusion}


We presented a graph neural network (GNN)-based simulator for modeling the uniaxial compression dynamics of two-dimensional disordered elastic networks (DENs). The model accurately reproduced system dynamics and emergent properties, such as the Poisson's ratio and its strain dependence. Despite being trained on only a small number of systems, the simulator demonstrated strong efficiency in learning system-to-property relationships.

Beyond replicating observed dynamics, a key finding of this work is the model’s ability to generalize well beyond its training distribution of Poisson's ratios. Even when the Poisson's ratio distribution in the training set was purposefully limited to a small range, the simulator still accurately predicted the behavior of systems with Poisson's ratios outside that range, a capability particularly valuable for design and exploration tasks. Additionally, the model demonstrated the ability to generalize across other factors, including system temperature and strain amplitude.
 
These results highlight the potential of using dynamical data to train machine learning models that are both system and data-efficient, and capable of extrapolating to system properties beyond the training domain. We anticipate that such approaches can help accelerate materials discovery and design, particularly in contexts where generating large datasets is costly or impractical.

\section{Supplementary Material}
\label{supp}
The supplementary material includes the simulation and data generation details (Sec. 1), illustrations of the complexity of dynamics (Sec. 2), and the behavior of the complexity coefficient (Sec. 3).



\begin{acknowledgments}
The authors acknowledge support from the Israel Science Foundation (ISF) under grant number 1181/24.
\end{acknowledgments}
\section*{Data Availability}
Data used in this work is available upon request. 

\section*{References} 

\bibliography{references}

\end{document}


\title{Supporting Materials for Dynamical Data for More Efficient and Generalizable Learning: A Case Study in Disordered Elastic Networks} 



\author{Salman N. Salman}
\thanks{These authors contributed equally to this work.}
\affiliation{The Wolfson Department of Chemical Engineering, Technion -- Israel
Institute of Technology, Haifa 32000, Israel}

\author{Sergey A. Shteingolts}
\thanks{These authors contributed equally to this work.}
\affiliation{The Wolfson Department of Chemical Engineering, Technion -- Israel
Institute of Technology, Haifa 32000, Israel}

\author{Ron Levie}
\affiliation{Faculty of Mathematics, Technion -- Israel
Institute of Technology, Haifa 32000, Israel}

\author{Dan Mendels}
\email[E-mail: ]{danmendels@technion.ac.il}
\affiliation{The Wolfson Department of Chemical Engineering, Technion -- Israel
Institute of Technology, Haifa 32000, Israel}

\maketitle

\section{Details on training data generation}

\subsection{Molecular dynamics simulations}

\subsubsection{On unit conversion}

For our systems, the following fundamental quantities were chosen: The distance unit $\sigma$ was defined as an average bond length of all the DENs in the dataset, $m$ was the beads' mass and $\epsilon$ was set to be the average bond constant (see below). Hereinafter, all quantities are listed in these units.

\subsubsection{Elastic network generation}
Disordered elastic networks (DENs) were generated in a two-step process. \cite{Rocks2017DesigningNetworks} First, $N$ soft disks of four different types and diameters were placed randomly in equal proportions in a simulation box with periodic boundary conditions and allowed to relax to a local energy minimum using a standard jamming algorithm. \cite{Liu2010TheSolid} The frictional force between each pair of disks located within a contact distance $d=R_i + R_j$ is responsible for interaction:
\[F=(k_n\delta \textbf{n}_{ij} - m_{eff}\gamma_n \textbf{v}_n) - (k_t \Delta s_t + m_{eff} \gamma_t \textbf{v}_t),\]
where $\delta$ is an overlap distance of two disks, $\textbf{n}_{ij}$ is the unit vector along the line connecting the centers of two disks, $k_n$ and $k_t$ are elastic constants for normal and tangential contact, respectively, $\gamma_n$ and $\gamma_t$ are viscoelastic constants for normal and tangential contact, $m_{eff} = \frac{M_i M_j}{M_i + M_j}$ is an effective mass of two disks $i$ and $j$ with individual masses $M_i$ and $M_j$, $\textbf{v}_n$ and $\textbf{v}_t$ are the normal and tangential components of the relative velocity of the two disks, and $\Delta s_t$ is the tangential displacement vector between the two disks. In our cases, $k_t$, $\gamma_n$ and $\gamma_t$ were set to zero, while $k_n$ was set to 1.0, so the resulting force between the two disks is simply expressed as follows:
\[F=\delta \textbf{n}_{ij}\]

During the energy minimization, the system was slowly cooled from $T_{start}=5\times10^{-6}$ to $T_{end}=1\times10^{-6}$.
After that, the beads were placed at the center of each disk and harmonic bonds were created between those beads that were in contact with each other, with the contact distance defined as the sum of disks' radii $d_{ij} = R_i + R_j$. Harmonic bonds were assigned an elastic energy in the form:
\[E(r_{ij}) = K_b (l - l_{re})^2, \]
where $K_b$ is the bond coefficient with units $energy/distance^2$, $l_{re}$ is the bond rest length, and $l$ is the distance between the two beads. For each bond, the value of $K_b$ was set to $1/l_{re}$. The resulting network configuration was then scanned for "dangling" beads, i.e., those with less than 3 connections to their neighbors, and those dangling beads were removed from the network. This procedure was applied iteratively until no such beads were left. To include angle-bending constraints, for each pair of bonds sharing one common bead an angle with an harmonic potential was assigned, with the energy in the form:
\[E(\theta_{km}) = K_{ang} (\theta_{km} - \theta_{km}^0)^2, \]
where $K_{ang}$ is the angle coefficients for angle $\theta_{km}$ between the pair of edges $k$ and $m$, and $\theta_{km}^0$ is the rest angle. For the final dataset, $K_{ang}$ was set to $0$ which further enhanced the networks' dynamics complexity and the model's ability to generalize.

\subsubsection{Uniaxial compression simulation}
Compression simulations of the 2d DENs were carried out using LAMMPS \cite{Thompson2022LAMMPSScales} molecular dynamics simulator. The systems were simulated within the isenthalpic ensemble with constant temperature $T=1\times 10^{-10}$ , $k_B = 1$ , damping parameter of $7.1 \times 10^{-3}$ and time step $\tau = 7.13 \times 10^{-6}$ controlled by Langevin thermostat.\cite{Schneider1978Molecular-dynamicsTransitions} The compression was simulated using the \textbf{fix deform} command of LAMMPS with a constant strain rate $s_r = 1\times10^{-8}$ and the maximum strain not exceeding 5\% of the initial box dimension along the compressed axis. Constant zero-pressure along the non-compressed axis was applied on the simulation box perpendicular to the direction of compression using the Parrinello-Rahman barostat, \cite{Parrinello1981PolymorphicMethod} which allowed the simulation box to change volume based on the mechanical properties of the network.

\subsubsection{Generating networks with low Poisson values}

Networks with low Poisson values were generated by optimizing previously generated high Poisson value networks (obtained with the methodology described above) using the global node optimization strategy\cite{Shen2024AnDensity} implemented in Python 3.2.12 using a custom-written code.  The algorithm is based on gradient descent and aims to minimize Poisson's ratio $\nu$ by altering individual node positions. At each iteration $n$, the coordinate of each node $i$, $\textbf{r}_i$ is displaced one by one by $5 \times 10^{-4}$ in an axis $\eta \in \{x,y\}$, the bond lengths and angles are recalculated accordingly. Then, $\nu$ for the resulting network configuration is measured (see section \ref{elastic_mod}) and the gradient is computed: 
\[\frac{\partial(\nu + \mathcal{L}_1 + \mathcal{L}_2)}{\partial r_{i\eta}},\]
where $r_{i\eta}$ is the displacement of node $i$ in the axis $\eta$, while $\mathcal{L}_1$ and $\mathcal{L}_2$ are the internode distance and angle constraints, respectively, which are defined as follows:
\begin{eqnarray*}
\mathcal{L}_1 &=& 0.1 \sum_{i, j} H(r_{min} - r_{ij})(r_{ij} - r_{min})^2, \\
\mathcal{L}_2 &=& 0.01 \sum_{j} H(\theta_{min} - \theta_{j})(\theta_{j} - \theta_{min})^6, \\
\end{eqnarray*}
where $H$ is the Heaviside step function, while $r_{min}$ and $\theta_{min}$ are the minimal values for bond distances and angles, respectively. The latter were chosen by trial and error, with the final values set to 0.2 and 15$^{\circ}$. To prevent structure from changing too aggressively and make the optimization process more stable, gradients were clipped to 0.01 by absolute value. Finally, the initial node positions are updated using the $N\times2$ gradient matrix obtained at the previous step:
\[r_{i\eta}^{n+1} = r_{i\eta}^{n} + \lambda \frac{\partial(\nu + \mathcal{L}_1 + \mathcal{L}_2)}{\partial r_{i\eta}},\]
where $\lambda$ is the learning rate, which was set to 4. The process is repeated iteratively until either \textit{a}) the desired target value for $\nu$ is reached or \textit{b}) the number of iterations exceeds the maximum of 50.

\subsubsection{Elastic moduli}
\label{elastic_mod}
Elastic moduli were computed using LAMMPS molecular dynamics simulator. For each trial bond removal, a compression simulation with a small strain $\epsilon = 1 \times 10^{-4}$ was performed and corresponding elastic constants $C_{ij}$ were calculated. Notably, as described in \cite{Reid2018}, bead-spring networks lose rigidity below coordination number Z=4, which necessitates the addition of angle bending coefficients. For two-dimensional isotropic systems, such as the 2d disordered elastic networks considered here, the elastic tensor is represented by a $3\times3$ matrix, which contains three independent elastic constants $C_{ij}$ due to the symmetry constraints of an isotropic material: $C_{11}$, $C_{12}$ and $C_{66}$. These relate the stress along $i$ axis to the strain along $j$ axis as follows: $\sigma_i = C_{ij} \cdot \epsilon_j$. Since the studied networks were not completely isotropic, as a precaution elastic constants $C_{11}$ and $C_{12}$ were calculated as averages, e.g., $C_{11}^a = (C_{11} + C_{22})/2$. From this, the desired elastic moduli were calculated as follows:
\begin{eqnarray*}
B &=& (C_{11}^a + C_{12}^a)/2 \\
G &=& (C_{11}^a - C_{12}^a)/2 \\
\nu &=& \frac{(1 - G/B)}{(1+G/B)}
\end{eqnarray*}

\subsubsection{Poisson's ratio estimation}
To evaluate the model's predictive power, we estimate the Poisson's ratio for the model rollout using the following expression:
\[\nu_{est} = -D_y/D_x,\]
where $D$ is the approximation for simulation box change at time t with respect to time 0:
\[D_{\eta} = l_{\eta}^0 \times \frac{\sum_{i=1}^N |r^{t}_{\eta}|}{\sum_{i=1}^N |r^0_{\eta}|} -l_{\eta}^0,\] 
where $l_{\eta}^0$ is the simulation box size in the $\eta$ dimension at the beginning of compression, 
$|r^{0}_{\eta}|$ and $|r^{t}_{\eta}|$  the distances between each node $i$ and the simulation box center (0 , 0) at the first and the last rollout steps, respectively, and N the number of nodes in the network. While not strictly accurate, this procedure provides a good enough estimation, with errors not exceeding 5\%. 

\newpage
\section{Dynamics complexity}

\begin{figure}[H]
    \centering
    \includegraphics[width=6in]{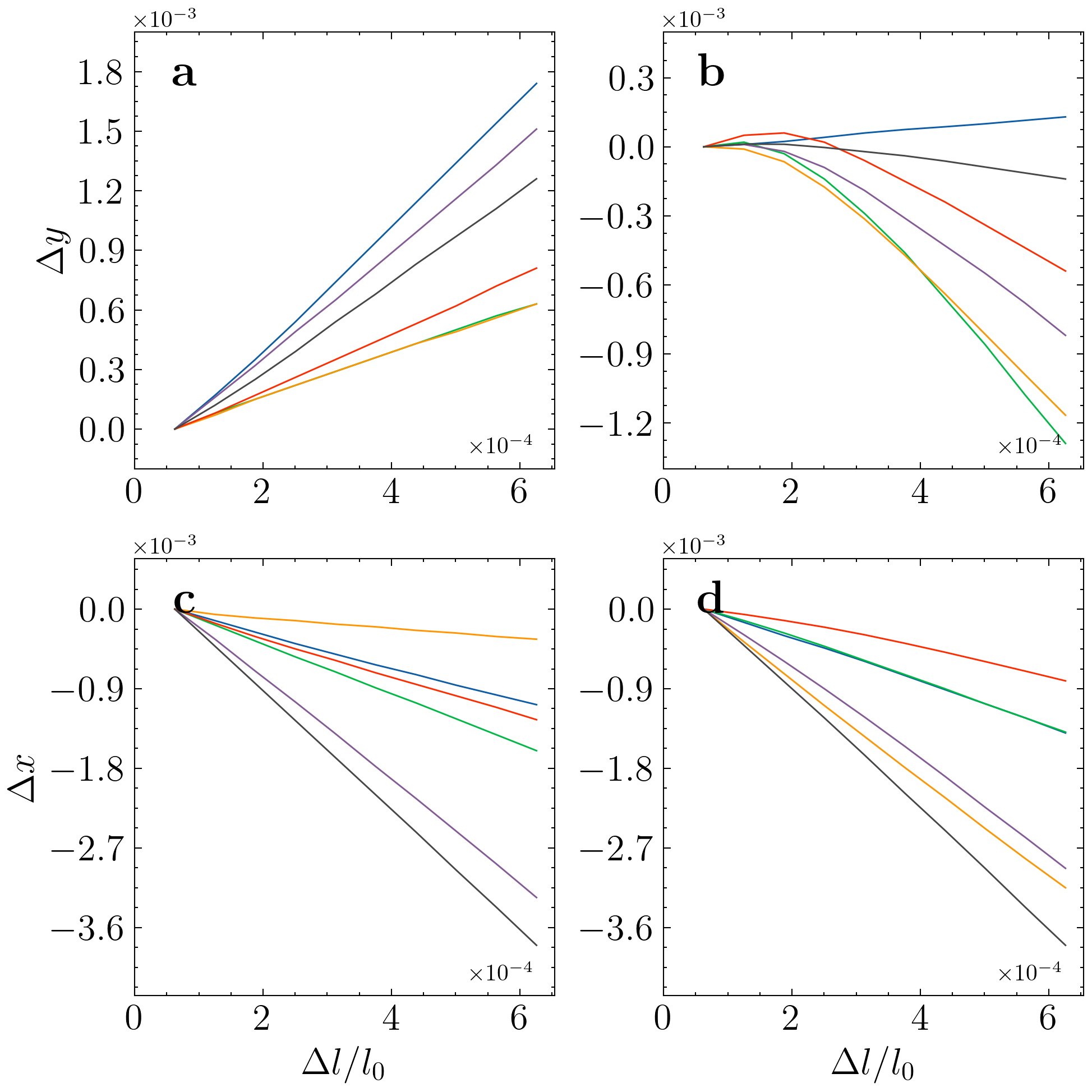}
    \caption{Node displacement in $x$ (c,d) and $y$ (a,b) for an auxetic (b,d) and a non-auxetic (a,c) simulation}
    \label{fig:node-trj}
\end{figure}

\newpage
\section{Complexity over time}

\begin{figure}[H]
    \centering
    \includegraphics[width=6in]{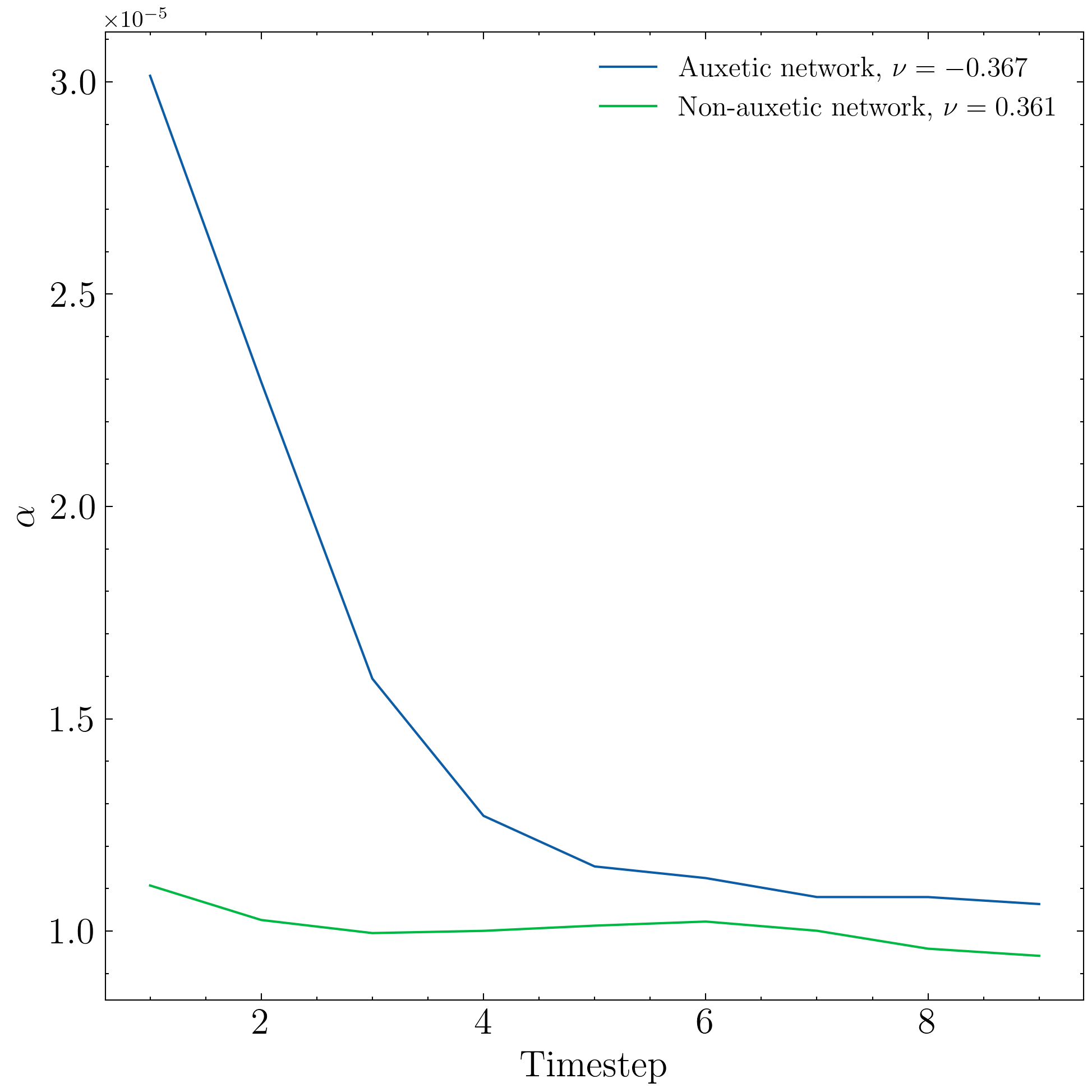}
    \caption{Behavior of $\alpha$ for auxetic vs non-auxetic simulations}
    \label{fig:alpha}
\end{figure}

We calculated the complexity coefficient $\alpha =  \frac{1}{NL}\sum_{i = 0}^{N} \sum_{t=t_0}^{L} |\textbf{v}_{t+1,i} -\textbf{v}_{t,i}|$ for two simulations, an auxetic and non-auxetic one, and plotted the behavior of $\alpha$. As Fig.\ref{fig:alpha} shows, auxetic simulations have a peak at the beginning that gradually drops and converges to a similar $\alpha$  value for both simulations. 

\newpage
\section{Model node dynamics predictions at high temperature}

\begin{figure}[H]
    \centering
    \includegraphics[width=6in]{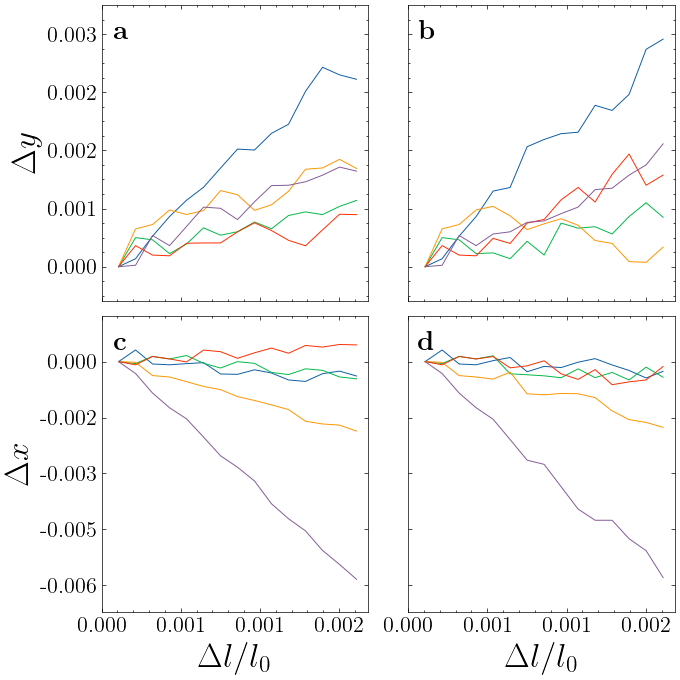}
    \caption{Node trajectories for a model trained  and tested on data simulated at $T=10^{-7}$}
    \label{fig:node-trj-high-T}
\end{figure}

\newpage
\bibliography{references}